\documentclass[12pt]{article}
\usepackage{amsfonts}
\usepackage{amssymb}
\usepackage{amsthm}
\usepackage{hyperref}
\usepackage{amssymb,amsmath,amsthm}
\usepackage{enumerate}
\usepackage{graphicx}

\newcommand{\sss}{\setcounter{equation}{0}}
\newtheorem{theorem}{THEOREM}[section]

\newtheorem{lemma}[theorem]{LEMMA}

\newtheorem{corollary}[theorem]{COROLLARY}

\newtheorem{remark}[theorem]{REMARK}

\newtheorem{definition}[theorem]{DEFINITION}

\newcommand{\ere}{ {\mathbb R}}
\def\cl{{\mathbb R}_0^3}

\newcommand{\CE}{{\mathbb C}}
\def\beq{\begin{equation}}
\def\ene{\end{equation}}

\newcommand{\bull}{\hfill $\Box$}

\def\qed{\ifhmode\unskip\nobreak\fi\ifmmode\ifinner
\else\hskip5pt\fi\fi\hbox{\hskip5pt\vrule width4pt height6pt
depth1.5pt\hskip1pt}}

\def\var{\varepsilon}

\def\e{\mathbf E}
\def \b{\mathbf B}
\def\d{\mathbf D}
\def\h{\mathbf H}
\def\x{\mathbf x}
\def\y{\mathbf y}
\def\c{\mathbf c}
\def\H{\mathcal H}
\def\C{\mathbf C}

\def\i{\hbox{\rm in}}
\def\s{\hbox{\rm sc}}
\begin{document}
\baselineskip=20 pt
\parskip 6 pt

\title{A Rigorous   Analysis  of High Order
Electromagnetic Invisibility  Cloaks
\thanks{ PACS classification scheme 2006: 41.20.Jb, 02.30.Tb,02.30.Zz, 02.60.Lj. AMS 2000 classification
35L45, 35L50, 35L80, 35P25, 35Q60, 78A25, 78A45.} \thanks{ Research partially
supported by  CONACYT under Project P42553­F.}}
 \author{  Ricardo Weder\thanks{ †.On leave of absence from Departamento de M\'etodos
 Matem\'aticos  y Num\'ericos. Instituto de Investigaciones en Matem\'aticas Aplicadas y en Sistemas.
 Universidad Nacional Aut\'onoma de
M\'exico. Apartado Postal 20-726, M\'exico DF 01000. Fellow Sistema Nacional de Investigadores.}
\\Department of Mathematics and Statistics \\
University of Helsinki \\
P.O. Box 68 (Gustaf Hallstromin katu 2b)
FI-00014. Finland
\\ weder@servidor.unam.mx}

\date{}
\maketitle
\begin{center}
\begin{minipage}{165mm}
\centerline{{\bf Abstract}}
\bigskip
There is currently a great deal of interest in the invisibility cloaks recently proposed by Pendry et al.
that are based in the transformation approach. They obtained their results using first order transformations.
 In  recent papers Hendi et al. and Cai et al. considered  invisibility cloaks with high order transformations.
In this paper we study high order electromagnetic invisibility
cloaks in transformation media obtained by
 high order transformations  from  general anisotropic media. We consider the case where there
is a finite number of spherical cloaks located in different points
in space. We prove that  for any incident plane wave, at any
frequency, the scattered wave is identically zero. We also consider
the scattering of finite energy wave packets. We prove that the
scattering matrix is the identity, i.e., that for any incoming wave
packet the outgoing wave packet is the same as the incoming one.
This proves that the invisibility cloaks can not be detected in any
scattering experiment with electromagnetic waves in
 high order transformation media, and in particular in the first order transformation media
of Pendry et al. We also prove that the  high order invisibility
cloaks, as well as the first order ones, cloak passive and active
devices.  The cloaked objects completely decouple from the exterior.
Actually, the cloaking outside is independent of what is inside the
cloaked objects. The electromagnetic waves inside the cloaked
objects can not leave the concealed regions and viceversa, the
electromagnetic waves outside  the cloaked objects can not go inside
the concealed regions.
\end{minipage}
\begin{minipage}{165mm}
As we prove our results for media that are obtained by transformation from general anisotropic materials, we
prove  that it is possible to cloak objects inside general crystals.
\end{minipage}
\end{center}


\section{Introduction}\sss

Recently \cite{pss1} considered electromagnetic invisibility cloaks based in the transformation
method that offer the theoretical and practical possibility of hiding objects from observation by
electromagnetic waves. The results in \cite{pss1} were obtained in the geometrical optics approximation.
Numerical simulations were reported in \cite{ccks} and \cite{cpss} and a  experimental verification of cloaking
was given in \cite{smjcpss}. In the case of one spherically symmetric cloak \cite{cwzk}   proved by an explicit
calculation in spherical coordinates that the scattered wave is identically zero at each frequency and that
the electromagnetic waves outside the spherical cloaked object can not enter the concealed region.
These papers considered first order transformations.
 The recent papers \cite{hen, ho}  consider  high order transformations that offer new
possibilities. In all these studies the transformation media were
obtained by transformations from isotropic media.

In this paper we consider electromagnetic invisibility cloaks in high order transformation media that are
obtained by  high order transformations  from  general anisotropic media. Moreover, we assume that there
 are several cloaked objects located in different points in space.

 We prove that  for any incident plane wave, at any frequency , the scattered wave is identically zero.
This generalizes, with a different proof, the results obtained in \cite{cwzk}. Note that in our case
separation of variables can not be used as there is no symmetry since we have a finite number of cloaks
in different points in space and also because we transform from  general anisotropic media.

We also consider the scattering of finite energy wave packets. We prove that the scattering matrix is the
identity, i.e., that for any incoming wave packet  the outgoing wave packet is the same as the incoming one.

These results  prove that the invisibility cloaks can not be detected in any scattering experiment with
electromagnetic waves in  high order transformation media, and in particular, in the first
order transformation media of \cite{pss1}.

We also prove that the  high order invisibility cloaks, as well as the first order ones, cloak passive and
active devices.  The cloaked objects completely decouple from the exterior. Actually, the cloaking
outside is independent of what is inside the cloaked objects. The electromagnetic waves inside
the cloaked objects can not leave the concealed regions and viceversa, the electromagnetic waves outside
the cloaked objects can not go inside the concealed regions.

In  \cite{we3} we considered first order transformation media. Here we present
the generalization to high order transformations, and we add new results.

The fact that the electromagnetic waves inside the cloaked objects can not leave the concealed region, that we
had already proved in \cite{we3}   in the case of first order transformations, has recently been verified in
the case of one spherical cloak in a first order transformation medium, by an explicit computation in
spherical coordinates,  by \cite{zcwk}, where also
a  physical mechanism based in surface voltages is presented to physically explain why the electromagnetic
waves cannot leave the concealed region.

Our results are based in Von Neumann's method of self-adjoint extensions. This is a very powerful technique
that allows us to settle in an  unambiguous way the mathematical problems posed by the singularities
of the inverse of the permittivity and the permeability of the transformation media in the boundary
of the cloaked objects. It also allows us to identify the appropriate boundary condition when cloaking is
formulated as a boundary value problem. Namely, that the tangential components of the electric and the magnetic
fields have to vanish at the outside of the boundary of the cloaked objects. See Remark \ref{def-2.4}. We
have already proven this result for first order transformations in \cite{we3}.  This boundary condition is
self-adjoint in our case because the permittivity and the permeability are degenerate at the boundary of the
cloaked objects.

As we prove our results for media that are obtained by transformation from general anisotropic materials, we
prove  that it is possible to cloak objects inside general crystals.

As it is often the case in the papers on electromagnetic invisibility cloaks, I make the assumption that the
media are not dispersive. This is a widely used idealization. As is well known, metamaterials are dispersive,
and, furthermore, when the the permittivity and the permeability have eigenvalues less than one, dispersion
comes into play in order that the group velocity does not exceeds the speed of light. This idealization means
that we have to take a narrow enough range of frequencies in order that we can analyse the cloaking  effect
without taking dispersion into account. In practice this means that cloaking will only be approximate.

For a related method for cloaking in two dimensions see \cite{le}. For earlier results in cloaking for
conductivity problems see \cite{bm}, the references quoted there, and \cite{glu1,glu2} .
In \cite{gklu}  cloaking is studied in the context of the
Dirichlet to Neumann map. For a cylindrical invisibility cloak with first order transformation see
\cite{zcwrk}. See also \cite{lp} and \cite{sps} for other related results

The paper is organized as follows. In Section 2 we introduce our formalism, we give the definition of solutions
with locally finite energy and we  obtain the cloaking boundary condition. Moreover,  we prove that
the electromagnetic waves inside the cloaked objects can not go outside  and viceversa.  In Section 3 we prove that the scattered waves
are zero for all frequencies  and all incoming plane waves. In Section 4 we prove that the scattering matrix is
the identity for all incoming wave packets. In Section 5 we give the proof of Theorems \ref{th-2.3}, and
 \ref{th-2.5}. In Section 6 we discuss generalizations of our results. Finally, we give a brief conclusion
 and  outlook where we also comment on  cloaking objects inside general anisotropic media, in particular inside
 general crystals.

\section{Electromagnetic Cloaking}
\sss
Let us consider Maxwell's equations in $\ere^3$, in the time domain,

\begin{eqnarray}
\nabla \times \e &=& -\frac{\partial}{\partial t}\b, \,\, \nabla \times \h \,=\,\frac{\partial }{\partial t} \d,
\label{2.1} \\\nonumber\\
 \nabla \cdot \b&=&0, \nabla \cdot \d\,=
 \,0,
\label{2.2}
 \end{eqnarray}
and in the frequency domain, assuming a periodic time dependence of  $\e,\h$  given by \linebreak
$e^{-iw t}$, with $\omega$
the frequency,
\begin{eqnarray}
\nabla \times \e &=&  i\omega \b, \,\, \nabla \times \h \,=\,-i\omega \d,\,\, \omega \neq 0,
 \label{2.3}\\
 \nabla \cdot \b&=&0, \nabla \cdot \d\,=
 \,0,
\label{2.4}
 \end{eqnarray}
where we have suppressed the factor $e^{-i\omega t}$ in both sides. Note that (\ref{2.4})
follows from (\ref{2.3}).

We study the propagation of electromagnetic waves -that  satisfy  Maxwell's equation- in the case where there
is a finite number of high-order spherical invisibility  cloaks located in different points in space.

For simplicity let us first consider the case where there is only one cloak located at $\x =0$.
See Figure \ref{fig1}. We designate the Cartesian coordinates of $\x$ by $x^\lambda, \lambda=1,2,3$. To define
 the  high-order
transformation media we introduce another copy of $\ere^3$, denoted by $\cl$. The points in $\cl$ are denoted
by $\y$ with coordinates
$y^\lambda, \lambda=1,2,3$. We designate by $\hat{\x}= \x/|\x|, \hat{\y}= \y/|\y|$ unit vectors.
 Consider the following transformation from
$\cl\setminus \{0\}$ to $\ere^3$ \cite{ho,pss1},

\beq
\x=\x(\y)=f(\y):= g(|\y|) \hat{\y}.
\label{2.5}
\ene
In spherical coordinates this transformation changes the radial coordinate but leaves the angular
coordinates constant, i.e., $|\x|= g(|\y|), \hat{\x}=\hat{\y}$. Given $ 0 <a < b$  we wish that this
transformation sends the punctuated ball $ 0 < |\y| \leq b$ onto the concentric shell $ a < |\x| \leq b$,
that it is the
identity for $|\y| \geq b$ and that it is one-to-one. Then, we assume that  $g$ satisfies the following
conditions.

\begin{definition}\label{def-2.1}
For any positive numbers $a, b $ with $ 0 <a < b$,  we say the $g$ is a cloaking function  in
$[0,b]$ if  $g(\rho)$ is twice continuously
differentiable on $[0,b],  g(0)=a, g(b )=b$,  and
$g'(\rho):=\frac{d}{d\rho}g(\rho)> 0, \rho \in [0,b]$.
\end{definition}
These high order transformations were introduced in \cite{ho} for the case of a cylindrical cloak. They
imposed the condition $g'(b)=1$. Since for our spherical cloaks we do not need this condition  we do not
assume it.
We define,

\beq \begin{array}{c}
\x=\x(\y)=f(\y):= g(|\y|) \hat{\y},\, \hbox{\rm for}\,0< |\y|\leq b, \\
\x=\x(\y):= \y,\, \hbox{\rm for}\, |\y| \geq b.
\end{array}
\label{2.6}
\ene

With these conditions (\ref{2.6}) is a bijection from $\cl \setminus\{0\}$
onto $\ere^3 \setminus B_a(0)$, where by

\beq
B_r(\x_0):= \left\{ \x \in \ere^3: |\ x- \x_0| \leq r \right\}
\label{2.7}
\ene
we denote the closed ball  of center $\x_0$ and radius $r$.
Moreover, it blows up the point $0$ onto the sphere
$|\x|=a$. It sends the punctuated ball   $ 0 < |\y| \leq b$ onto the concentric shell $ a <|\x | \leq b$
and  it is the identity for $ |\y| \geq b$.
 It is twice continuously  differentiable away from the sphere $|\y|=b$, where it can have
 discontinuities in the derivatives depending on the values of the derivatives of $g$ at $b$.

 In \cite{ho}  the quadratic case
$$
g(\rho)= \left[ 1-\frac{a}{b}+ p(\rho -b)\right] \rho
+a
$$
 with $ p \in \ere$  was discussed in connection with a cylindrical cloak in an approximate
 transformation medium.
 In \cite{pss1} the first order  case $ g(\rho)= \frac{b-a}{b}\rho +a$  was considered.

The closed ball $K:=\{ \x \in \ere^3: |\x|\leq a\}$ is the region that we wish to conceal, and we call it the
cloaked object. The  spherical shell $a < |\x| \leq b$ is the cloaking layer. The union of the
cloaked object  and the cloaking layer is the spherical cloak, that in this case is just the closed
ball of center zero and
radius $b$. The domain  $ |\x| > b$ is the exterior of the spherical cloak.

To have several cloaks we just put a finite number of these spherical cloaks in different points in space
at a finite positive  distance from each other, in order that they do not intersect. See Figure \ref{fig2}.
Let us take  as
centers of the cloaks points  $ \mathbf{c}_j \in \ere^3, j=1,2,\cdots N$ where $N$ is the number of cloaks
 and
${\mathbf c}_j \neq {\mathbf c}_l, j \neq l,  1\leq j,l  \leq N$. We take $ 0 < a_j < b_j,
 \,\hbox{\rm and cloaking functions}\, g_j\, $  that satisfy the conditions of Definition
 \ref{def-2.1} for $a_j,b_j, j=1,2,3\cdots N$,
and we define the following transformation from $ \cl\setminus \{\c_1,\c_2,\cdots , c_N\}$ to $\ere^3$.

\beq
\begin{array}{c}
 \x=\x(\y) = f(\y):= \c_j + g_j(|\y-\c_j|) \,\, \widehat{\y-\c_j}, \y \in B_{b_j }({\mathbf c}_j),
\, j=1,2, \cdots, N, \\\\
 \x=\x(\y) = f(\y):=\y , \y \in  \cl \setminus \cup_{j=1}^N B_{b_j }({\mathbf c}_j),
\end{array}
\label{2.8}
\ene
where   $B_{b_j }({\mathbf c}_j)$ are balls in $\cl$.

The cloaked objects that we wish to conceal are  given by,

\beq
K_j:=\left\{ \x  \in \ere^3 : |\x-\c_j |\leq   a_j \right\}, j=1,2,\cdots,N.
\label{2.9}
\ene
The concentric spherical shells  $a_j < |\x- \c_j| \leq b_j, j=1,2,\cdots,N$
 are  the cloaking layers.  The spherical cloaks are  the balls $B_{b_j}(\c_j)$ in $\ere^3$. We denote by
$K$  the union of all the cloaked objects,
\beq
K:= \cup_{j=1}^N K_j.
\label{2.10}
\ene
The domain
\beq
\ere^3 \setminus \cup_{j=1}^N B_{b_j}(\c_j)
\label{2.11}
\ene
is the exterior of the all the spherical cloaks.
We assume that   the spherical cloaks are at a positive distance of each other,

$$
\hbox{min distance}\left( B_{b_j}(\c_j),  B_{b_l}(\c_l) \right) >0, j\neq l, j,l =1,2,\cdots, N.
$$
Denote
$$
\Omega_0:= \cl \setminus \{\c_1, \c_2, \cdots,\c_N\},\,\, \Omega:= \ere^3 \setminus K.
$$

 Then, (\ref{2.8}) is a    bijection from $\Omega_0$ onto
$\Omega$,   and  for $j=1,2, \cdots,N$ it blows up the point $\c_j$ onto the sphere
$|\x- \c_j|=a_j$. It sends the punctuated ball $ 0 < |\y-\c_j| \leq b_j$ onto the concentric
shell $ a_j <|\x -\c_j| \leq b_j$
and  it is the identity for $ \y \in  \cl \setminus \hbox{\rm interior}\left(\cup_{j=1}^N B_{b_j }
(\c_j)\right)$. It is twice continuously differentiable away from the spheres $|\y-\c_j|=b_j$, where it
can have
 discontinuities in the derivatives depending on the values of the derivatives of $g_j$ at $b_j$.

For any open set $O$ and for any $ n=1,2, \cdots$,  let us denote by $C^n(O)$ the set of all $\CE -$valued
functions that are continuous together with all
its derivatives of  order up to $n$, and  by $C^n_0(O)$ the functions in $C^n(O)$ that have compact support in
$ O$, i.e. such that the closure of the set of points where they are different from zero is
bounded in $O$. In other words, the closure of set of points where the functions
 are different from zero is bounded, and they are zero in a neighborhood of the boundary of $O$. By $C(O),
 C_0(O)$, we denote, respectively, the continuous functions in $O$ and the continuous functions with compact
 support in $O$.

 We denote  the elements of the Jacobian matrix by
$ A^\lambda_{\lambda'}$,
\beq
A^\lambda_{\lambda'}:= \frac{\partial x^\lambda}{\partial y^{\lambda'}}.
\label{2.12}
\ene
 Note that
$A^\lambda_{\lambda'}\in
 \C^1\left(\Omega_0 \setminus \cup_{j=1}^N \partial B_{b_j}(\c_j)\right)$, and that it can have discontinuities
 on $\cup_{j=1}^N \partial B_{b_j}(\c_j)$ depending on the derivatives of $g_j$ at $b_j$.
 We designate by $A^{\lambda'}_\lambda$ the elements
of the Jacobian of the inverse bijection,
$\y=\y(\x)=f^{-1}(\x)$,

\beq
A^{\lambda'}_{\lambda}:= \frac{\partial y^{\lambda'}}{\partial x^{\lambda}}.
\label{2.13}
\ene
$A^{\lambda '}_{\lambda}\in
 \C^1\left(\Omega \setminus \cup_{j=1}^N \partial B_{b_j}(\c_j)\right)$, and it can have discontinuities
 on $\cup_{j=1}^N \partial B_{b_j}(\c_j)$ depending on the derivatives of $g_j$ at $b_j$. Let us denote
 by $\Delta$ the determinant of the Jacobian matrix
(\ref{2.12}). See (\ref{5.2}) for the calculation of $\Delta$ in
closed form. Note that it diverges at $\partial K= \partial
\Omega$.

We take here the  {\it material interpretation} and we consider our
transformation as a bijection between two different spaces,
$\Omega_0$ and $\Omega$. However, our transformation can be
considered, as well, as a change of coordinates in $\Omega_0$. Of
course, these two points of view are mathematically equivalent. This
means, in particular, that under our transformation  Maxwell's
equations in $\Omega_0$ and in $\Omega$ have the same
invariance  that they have under change of coordinates in
three-space. See, for example, \cite{po}. Let us denote by
$\e_0,\h_0,\b_0, \d_0,  \varepsilon^{\lambda\nu}_0,
\mu^{\lambda\nu}_0$, respectively, the electric and magnetic fields,
the magnetic induction, the electric displacement, and the
permittivity and permeability of $\Omega_0$.  The, $
\varepsilon^{\lambda\nu}_0, \mu^{\lambda\nu}_0$, are positive,
hermitian matrices that are constant in $\Omega_0$.

The electric field is a covariant vector that transforms as,

\beq E_\lambda(\x) = A_\lambda^{\lambda'}(\y)E_{0,\lambda'}(\y).
\label{2.15} \ene

The magnetic field $\h$ is a covariant pseudo-vector, but as we only
consider space transformations with positive determinant, it also
transforms as in (\ref{2.15}). The magnetic induction $\b$ and the
electric displacement $\d$ are contravariant vector densities of
weight one that transform as

\beq B^\lambda(\x) =  \left(\Delta (\y)\right)^{-1}
A_{\lambda'}^{\lambda}(\y) B^{\lambda'}_0(\y),
\label{2.16}
\ene
with the same transformation for $\d$. The permittivity and
permeability are contravariant tensor densities of weight one that
transform as,
\beq
\varepsilon^{\lambda\nu}(\x)=  \left(\Delta
(\y)\right)^{-1} A^{\lambda}_{\lambda'}(\y)\, A^{\nu}_{\nu'}(\y)\,
\varepsilon^{\lambda' \nu'}_0(\y),
\label{2.17}
\ene
with the same transformation for $ \mu^{\lambda\nu}$.  Maxwell's equations
(\ref{2.1}-\ref{2.4}) are the same in both spaces $\Omega$ and
$\Omega_0$. Let us denote by $\varepsilon_{\lambda \nu},
\mu_{\lambda \nu}, \varepsilon_{0\lambda \nu}, \mu_{0\lambda \nu}$,
respectively, the inverses of the corresponding permittivity and
permeability. They are covariant tensor densities of weight minus
one that transform as,

\beq \varepsilon_{\lambda\nu}(\x)=  \Delta (\y)
A^{\lambda'}_{\lambda}(\y)\, A^{\nu'}_{\nu}(\y)\,
\varepsilon_{0\lambda' \nu'}(\y),\, \mu_{\lambda\nu}(\x)=  \Delta
(\y) A^{\lambda'}_{\lambda}(\y)\, A^{\nu'}_{\nu}(\y)\,
\mu_{0\lambda' \nu'}(\y).
\label{2.18} \ene
 Note that
\beq
 \det \varepsilon^{\lambda\nu}=\Delta^{-1} \det \varepsilon^{\lambda\nu}_0,\,
 \det \mu^{\lambda\nu}=\Delta^{-1} \det \mu^{\lambda\nu}_0,\,
\label{2.19}
\ene

\beq
\det \varepsilon_{\lambda\nu}=\Delta \det
\varepsilon_{0\lambda\nu},\,  \det \mu_{\lambda\nu}=\Delta \det
\mu_{0\lambda\nu}.
\label{2.20}
\ene

 Then, by (\ref{2.19}, \ref{2.20}, \ref{5.2}) the matrices $ \var^{\lambda \nu}, \mu^{\lambda \nu}$ are
 degenerate at
 $\partial K$ and  the matrices  $ \var_{\lambda \nu}, \mu_{\lambda \nu}$ are singular at
 $\partial K$.

We face now the problem that as the $\varepsilon^{\lambda\nu}$ and $\mu^{\lambda\nu}$ are degenerate at
the boundary of the cloaked objects $K$  we have to make precise what do we mean by a solution to
Maxwell's equations. As we will see the standard  rules that we apply in non-degenerate situations do not
apply here. These type of problems are not unusual in  mathematical physics and there is a well established
method to deal with them. Namely, Von
Neumann's theory of self-adjoint extensions \cite{rsII, rsIII}. Let us first consider the problem in
$\Omega$. We write Maxwell's equations in Schr\"odinger form. For this purpose we denote  by
$ {\mathbf \var}$ and $ {\mathbf \mu}$, respectively, the matrices with entries
$\var_{\lambda\nu}$ and $\mu_{\lambda\nu}$.  Recall that $\left(\nabla\times \e\right)^\lambda=
s^{\lambda\nu\rho} \frac{\partial}{\partial x_\nu }E_\rho$, where $s^{\lambda\nu\rho}$ is the permutation
contravariant pseudo-density of weight $-1$
(see section 6 of chapter II of \cite{po}, where a different notation is used).

We  define the following formal differential operator,
\beq
a_\Omega \left(\begin{array}{c}\e\\ \h\end{array}\right)=i \left(\begin{array}{c}
{\mathbf\var} \nabla\times \h\\- {\mathbf \mu} \nabla\times \e\end{array}\right).
\label{2.21}
\ene
Here, as usual, we denote, $\var \nabla\times \h:= \var_{\lambda\nu} (\nabla \times \h)^\nu$, and
$\mu \nabla\times \e=\mu_{\lambda\nu}
(\nabla \times \e)^\nu$.

Equation (\ref{2.1}) is equivalent to,
\beq
i\frac{\partial}{\partial t}\left(\begin{array}{c}\e\\\h\end{array}\right)= a_\Omega\left(\begin{array}{c}\e\\\h\end{array}\right).
\label{2.22}
\ene

and equation (\ref{2.3}) is equivalent to
\beq
\omega\left(\begin{array}{c}\e\\\h\end{array}\right)= a_\Omega\left(\begin{array}{c}\e\\\h\end{array}\right).
\label{2.23}
\ene
Note that since the matrices $\epsilon, \mu$ are singular at $\partial \Omega$ the operator $a_\Omega$ has
coefficients that are singular  at $\partial \Omega$. This is the reason why we have to be careful when
defining the solutions.

We have to define  equation (\ref{2.21}) in an appropriate linear subspace of the Hilbert space of all
finite energy fields in $\Omega$, that we define now. We designate  by $\H_{\Omega E}$ the Hilbert space of
all measurable, $\CE^3-$ valued functions defined on $\Omega$  that are square integrable with the weight
$\var^{\lambda\nu}$ and the scalar product,
\beq
\left(\e^{(1)}, \e^{(2)}\right)_{\Omega E}:= \int_{\Omega}E^{(1)}_\lambda\,\var^{\lambda\nu}\,
\overline{E^{(2)}_\nu}\, d\x^3.
\label{2.24}
\ene
Moreover,  we denote  by $\H_{\Omega H}$ the Hilbert space of all
measurable, $\CE^3-$ valued functions defined on $\Omega$  that are square integrable with the weight
 $\mu^{\lambda\nu}$ and the scalar product,
\beq
\left(\h^{(1)}, \h^{(2)}\right)_{\Omega H}:= \int_{\Omega}H^{(1)}_\lambda\,\mu^{\lambda\nu}\,
\overline{H^{(2)}_\nu}\, d\x^3.
\label{2.25}
\ene

The Hilbert space of finite energy fields in $\Omega$ is the direct sum

\beq
\H_\Omega:= \H_{\Omega E}\oplus \H_{\Omega H}.
\label{2.26}
\ene

For any open set $O$ the spaces $\C(O), \C_0(O), \C^n(O), \C^n_0(O), n=1,2\cdots$   are defined as
the spaces $C(O), C_0(O), C^n(O), C^n_0(O), n=1,2\cdots$ but with $\CE^6-$ valued functions.

We first define $ a_\Omega$ in a nice set of functions where it makes sense, that we take as $\C^1_0(\Omega)$.
In physical terms this means that we start with the minimal assumption that Maxwell's equations are satisfied
in classical sense away from the boundary of $\Omega$.
 $a_\Omega$ with domain $D(a_\Omega):= \C^1 _0(\Omega)$ is a symmetric operator in $\H_\Omega$, i.e.
$ a_\Omega \subset a_\Omega^\ast$.
To construct a unitary dynamics that preserves energy we have to analyze the self-adjoint extensions of
$a_\Omega$, what in physical terms means that we have to make precise in what  sense Maxwell's equations are
solved up to $\partial \Omega$. In other words,  to construct finite-energy solutions of (\ref{2.22}),

$$
\left(\begin{array}{c}\e\\\h\end{array}\right)(t)\in \H_\Omega,
$$

with constant energy
$$
\left(\e(t), \e(t)\right)_{\Omega E}+ \left(\h(t), \h(t)\right)_{\Omega H}=
\left(\e(0), \e(0)\right)_{\Omega E}+ \left(\h(0), \h(0)\right)_{\Omega H} < \infty,
$$
we have to demand that the initial finite energy fields , $ (\e(0), \h(0))^T$  belong to the domain
of one of the self-adjoint extensions of $a_\Omega$.  The key issue is that $a_\Omega$ has only one
self-adjoint extension, i.e. it is essentially self-adjoint. Before we state this result in a precise way
in Theorem \ref{th-2.3} let us discuss  its physical consequences. Let us denote by $A_\Omega$
the unique self-adjoint extension.

We denote by $ \hbox{\rm kernel}\,  A_\Omega$ the null subspace of
$A_\Omega$, i.e.,

$$
 \hbox{\rm kernel}\,  A_\Omega:=\left\{ \left(\begin{array}{c}\e\\\h\end{array}\right)\in D(A_\Omega):
 A_\Omega \left(\begin{array}{c}\e \\\h\end{array}\right)=0     \right\},
 $$
 and by
 $$
  \H_{\Omega\perp}:=\left(\hbox{\rm kernel}\,  A_\Omega\right)^\perp
  $$

 the orthogonal complement in $\H_\Omega$
 of  $\hbox{\rm kernel}\,  A_\Omega$. Equations (\ref{2.2}) are satisfied for all times if and
 only if $ (\e(0), \h(0))^T \in \left(\hbox{\rm kernel}\,  A_\Omega\right)^\perp$.
 Moreover, the unique  finite energy solutions to (\ref{2.1}, \ref{2.2}) with constant energy are
 constructed as follows.

We take any
\beq
\left(\begin{array}{c}\e \\\h\end{array}\right)\in \H_{\Omega\perp}\cap D(A_\Omega)
\label{2.27}
\ene
and we  obtain the  finite energy solution to  Maxwell's equations (\ref{2.1}, \ref{2.2}) as

\beq
    \left(\begin{array}{c}\e \\\h\end{array}\right)(t) =e^{-it A_\Omega} \left(\begin{array}{c}\e \\\h
    \end{array}\right).
\label{2.28}
\ene
This is the unique finite energy solution with constant energy, and with initial value at $t=0$ given by
(\ref{2.27}). Note that
 as $e^{-itA_\Omega} \H_{\Omega\perp}\subset
\H_{\Omega\perp}$ equations (\ref{2.2}) are satisfied for all times
if they are satisfied at $t=0$. The unitary group $e^{-it A_\Omega}$
is defined via functional calculus, but we can think of it as just
the operator that gives us the unique solution.  We can consider
more general solutions by means of the scale of spaces associated
with $A_\Omega$, but we do not go into this direction here.

 Solutions to (\ref{2.3},\ref{2.4}) in general do not have finite energy  because they do not have enough
 decay
 at infinity to be square integrable over all $\Omega$. Then, we only require  that they are of locally
 finite energy in the sense that the  electric  and the magnetic fields are square integrable over
 every bounded subset of $\Omega$, respectively, with the weight
 $\var^{\lambda\nu}$, and $\mu^{\lambda\mu}$.
 Moreover, in order that the problem (\ref{2.3},\ref{2.4}) is well-posed -in the sense that it is
 self-adjoint-
 the solutions with locally finite energy have to be locally in the domain of the only self-adjoint extension
 of $a_\Omega$, that is to say,  they  have to be in the domain of $A_\Omega$ when multiplied by any continuously
 differentiable function with
support in a bounded subset of $\overline{\Omega}$. Hence, we define,

\begin{definition} \label{def-2.2} (Solutions with Locally Finite Energy)
We say that the fields $(\e,\h)^T$ are a solution to (\ref{2.3},\ref{2.4})
with locally finite energy in $\Omega$ if they solve (\ref{2.3},\ref{2.4})
in distribution sense in $\Omega$, if, furthermore,  for every
bounded set $O \subset \Omega$

\beq
\int_{O}E_\lambda\,\var^{\lambda\nu}\, \overline{E_\nu}\, d\x^3 +
\int_{O}H_\lambda\,\mu^{\lambda\nu}\, \overline{H_\nu}\, d\x^3 <
\infty,
\label{2.29}
\ene
and if for any  continuously differentiable function $\phi$ with
bounded support in  $\overline{\Omega}, \, \phi \,(\e,\h)^T \in D
(A_\Omega)$.
\end{definition}

Similarly, given any bounded set $ O \subset \Omega$  we say that the fields $(\e,\h)^T$ are a finite
energy
solution in $O$ if (\ref{2.29}) is satisfied, if  the $(\e,\h)^T$ are a solution to (\ref{2.3}, \ref{2.4})
 in
distribution sense in $O$ and if for any continuously differentiable function $\phi$ with support in
$\overline{O},\,\, \phi(\e,\h)^T \in D(A_\Omega)$.

To state  Theorem \ref{th-2.3}  and to further analyze our  problem we consider now Maxwell's
equations in $\cl$ and we define
 the Hilbert spaces of electric and magnetic fields with  finite energy.
 The  $\e_0,\h_0,\b_0, \d_0$, were defined in $\Omega_0$,
but since $\cl\setminus \Omega_0 =\{\c_j\}_{j=1}^N$ is of measure zero, we can consider them as defined
 in
$\cl$, what we do below.

We denote by $\H_{0E}$ the Hilbert space of all
measurable, square integrable, $\CE^3-$ \,valued functions defined on $\cl$ with the scalar product,
\beq
\left(\e^{(1)}_0, \e^{(2)}_0\right)_{0 E}:= \int_{\cl}E^{(1)}_{0\lambda} \, \var^{\lambda\nu}_0\,
\overline{ E^{(2)}_{0\nu}}\, d\y^3.
\label{2.30}
\ene
We similarly define the Hilbert space,$\H_{0H}$,  of all
measurable, square integrable, \, $\CE^3-$ valued functions defined on $\cl$ with the scalar product,
\beq
\left(\h^{(1)}_0, \h^{(2)}_0\right)_{0 H}:=  \int_{\cl}H^{(1)}_{0\lambda} \, \mu^{\lambda\nu}_0\,
\overline{H^{(2)}_{0\nu}}\, d\y^3.
\label{2.31}
\ene
The Hilbert space of finite energy fields in $\cl$ is the direct sum

\beq
\H_0:= \H_{0 E}\oplus \H_{0 H}.
\label{2.32}
\ene

We now write  Maxwell's equations in $\cl$ in  Schr\"odinger form. As before we denote by
$ \var_0$ and $\mathbf \mu_0$, respectively, the matrices with entries $\var_{0\lambda\nu}$
and $\mu_{0\lambda\nu}$. By $a_0$ we denote
 the
following formal differential operator,

\beq
a_0 \left(\begin{array}{c}\e_0\\ \h_0\end{array}\right)=i \left(\begin{array}{c}
\var_0 \nabla\times \h_0\\- \mu_0 \nabla\times \e_0\end{array}\right).
\label{2.33}
\ene

Then, equation (\ref{2.1}) in $\cl$  is equivalent to,
\beq
i\frac{\partial}{\partial t}\left(\begin{array}{c}\e_0\\\h_0\end{array}\right)= a_0\left(\begin{array}{c}
\e_0\\\h_0\end{array}\right).
\label{2.34}
\ene

Let us denote by $\C^\infty_0(\cl)$ the set of all $\C^6-$valued continuously differentiable functions
on $\cl$
that have compact support in $\cl$ . Then, $a_0$ with domain
$\C^\infty _0(\cl)$ is a symmetric operator in $\H_0$, i.e., $ a_0 \subset a_0^\ast$. Moreover, it is essentially
self-adjoint in $\H_0$, i.e.,
it has only one self-adjoint extension, that we denote by $A_0$. Its domain is given by,

\beq
D(A_0)=\left\{ \left(\begin{array}{c}\e_0\\\h_0\end{array}\right) : a_0 \left(\begin{array}{c}\e_0\\
\h_0\end{array}\right) \in \H_0\right\},
\label{2.35}
\ene
and,
\beq
A_0 \left(\begin{array}{c}\e_0 \\\h_0\end{array}\right)= a_0 \left(\begin{array}{c}\e_0\\
\h_0\end{array}\right),\,
\left(\begin{array}{c}\e_0\\\h_0\end{array}\right)
\in D(A_0),
\label{2.36}
\ene
where the derivatives in the right-hand sides of (\ref{2.35}, \ref{2.36}) are taken in distribution sense in $\cl$.
These results follow easily from the fact that -via the Fourier transform-  $a_0$ is unitarily equivalent
 to multiplication by a matrix valued function  that is symmetric with respect to the scalar product
 of $\H_0$. Moreover, it follows from explicit computation that the only eigenvalue of $A_0$ is zero, that
 it
 has infinite multiplicity, and that,

\beq
\H_{0\perp}:= \left(\hbox{\rm kernel}\, A_0\right)^\perp = \left\{\left(\begin{array}{c}\e_0\\
\h_0\end{array}\right)\in \H_0:
\frac{\partial}{\partial x_\lambda}
\var^{\lambda\nu}_0E_{0\nu}=0 ,  \frac{\partial}{\partial x_\lambda}
\mu^{\lambda\nu}_0H_{0\nu}=0\right\}.
\label{2.37}
\ene
Furthermore, $A_0$ has no singular-continuous spectrum and its absolutely-continuous spectrum is $\ere$.
See,
for example, \cite{we1,we2}.

Taking any
\beq
\left(\begin{array}{c}\e_0 \\\h_0\end{array}\right)\in \H_{0\perp}\cap D(A_0)
\label{2.38}
\ene
we obtain a finite energy solution to  Maxwell's equations (\ref{2.1}, \ref{2.2}) in $\cl$  as follows

\beq
    \left(\begin{array}{c}\e_0 \\\h_0\end{array}\right)(t) =e^{-it A_0} \left(\begin{array}{c}\e_0 \\
    \h_0\end{array}\right).
\label{2.39}
\ene
This is the unique finite energy solution with initial value at $t=0$ given by (\ref{2.38}).
Note that as $e^{-itA_0} \H_{0\perp}\subset \H_{0\perp}$
equations (\ref{2.2}) are satisfied for all times if they are satisfied at $t=0$.

We denote by $U_E$ the following unitary operator from $\H_{0 E}$ onto $\H_{\Omega E}$,

\beq
\left(U_E \e_0\right)_\lambda(\x): = A^{\lambda'}_\lambda  E_{0\lambda'}(\y),
\label{2.40}
\ene
and by $U_H$ the unitary operator from $\H_{0H}$ onto $\H_{\Omega H}$,

\beq
\left(U_H \h_0\right)_\lambda(\x): = A^{\lambda'}_\lambda  H_{0\lambda'}(\y).
\label{2.41}
\ene

Then,
\beq
U:= U_E \oplus U_H
\label{2.42}
\ene
is a unitary operator from $\H_0$ onto $\H_\Omega$.

We prove the following theorem in Section 5.
\begin{theorem} \label{th-2.3}
The operator $a_\Omega$ is essentially self-adjoint, and its unique self-adjoint extension, $A_\Omega$,
satisfies
\beq
A_\Omega = U\, A_0\, U^\ast.
\label{2.43}
\ene
Furthermore, $A_\Omega$  has no singular-continuous spectrum and its absolutely-continuous spectrum is
$\ere$. The only eigenvalue of $A_\Omega$ is zero and it has infinite multiplicity.
Moreover,

\beq
\H_{\Omega \perp}:= \left(\hbox{\rm kernel}\, A_\Omega\right)^\perp = \left\{\left(\begin{array}{c}\e\\\h
\end{array}\right)\in \H_\Omega:
\frac{\partial}{\partial x_\lambda}
\var^{\lambda\nu}E_{\nu}=0 ,  \frac{\partial}{\partial x_\lambda}
\mu^{\lambda\nu}H_{\nu}=0\right\}.
\label{2.44}
\ene
\end{theorem}

The facts that $a_\Omega$ is essentially self-adjoint and that its unique self-adjoint extension $A_\Omega$
is
unitarily equivalent to the generator $A_0$ of the homogeneous medium are strong statements.
They mean that the only possible unitary dynamics in $\Omega$ that preserves energy is  given by
 (\ref{2.28}) and that this dynamics is unitarily equivalent
to the free dynamics in $\cl$ given by (\ref{2.39}). In fact, $\partial \Omega$ acts like a horizon for
electromagnetic waves propagating in $\Omega$ in the sense that the dynamics is uniquely defined without any
need to consider  the cloaked objects $K=\cup_{j=1}^N K_j$.
As we will prove below this implies, in particular,  electromagnetic cloaking for all frequencies in the
strong sense that  the scattered wave is identically zero at each frequency and that the scattering
operator in the time domain is the identity.

Formulating cloaking as a boundary value problem in $\Omega$ is of independent interest. For this purpose we
introduce the following boundary condition.

\begin{remark}\label{def-2.4}{\rm ( The Cloaking Boundary Condition)

Let  $(\e,\h)^T$  be a solution with locally finite energy in $\Omega$. According to Definition \ref{def-2.2}
they have to be locally in the domain of $A_\Omega$. By (\ref{2.43}) this implies that,

\beq
\left(\begin{array}{c}\e\\ \h\end{array}\right)= U \left(\begin{array}{c}\e_0\\ \h_0\end{array}\right)
\label{2.45}
\ene
with $(\e_0,\h_0)^T$ locally in the domain of $A_0$, i.e., $(\e_0,\h_0)^T$ is in the domain of $A_0$
 when multiplied by any function in $C^1_0(\cl)$.
It follows from (\ref{2.45}), and (\ref{5.1}) that the solutions with locally finite energy have to
satisfy the cloaking boundary condition,

\beq
\e\times \mathbf n=0, \h\times  \mathbf n=0, \,\, \hbox {\rm in }\,\, \partial \Omega=\partial K_+,
\label{2.46}
\ene

where $\partial K_+$ is the outside of the boundary of the cloaked objects and $\mathbf n$ is the normal
vector to $\partial K_+$.}
\end{remark}

Note that as $A_\Omega$ is the only self-adjoint extension of $a_\Omega$,  this is the only possible
self-adjoint boundary condition on $\partial K_+$. It is self-adjoint because the matrices
$\var, \mu$  are singular at $\partial K_+$. Then, cloaking as boundary value problem consists of
 finding
a solution to (\ref{2.3}, \ref{2.4}) in $\Omega$  with locally-finite energy
that satisfies the cloaking boundary condition  given in (\ref{2.46}).

\bull

Let us now consider  the propagation of electromagnetic  waves in the cloaked objects.
For this purpose we assume that in each $K_j$ the permittivity and the permeability are given
by $\var^{\lambda\nu}_j, \mu^{\lambda\nu}_j$, with
inverses $\var_{j\lambda\nu}, \mu_{j\lambda\nu}$ and where $\var_j, \mu_j$ are the matrices with entries
$\var_{j\lambda\nu}, \mu_{j\lambda\nu}$.
Furthermore, we assume that $  0 < \var^{\lambda\nu}(\x), \mu^{\lambda \nu}(\x) \leq C, \x \in K_j$ and that
for any compact set $Q$ contained in the interior
of $K_j$ there is a positive constant $C_Q$ such that $\det \var^{\lambda\nu}(\x) > C_Q,
\det\mu^{\lambda \nu}(\x) > C_Q, \x \in Q$. In other words, we only
allow for possible singularities of $\var_j, \mu_j$ on the boundary of $K_j$.

We designate  by $\H_{j E}$ the Hilbert space of all
measurable, $\CE^3-$ valued functions defined on $K_j$  that are square integrable with the weight
 $\var^{\lambda\nu}_j$ and the scalar product,
\beq
\left(\e^{(1)}_j, \e^{(2)}_j\right)_{j E}:= \int_{K_j}E^{(1)}_{j\lambda}\,\var^{\lambda\nu}_j\,
\overline{E^{(2)}_{j\nu}}\, d\x^3.
\label{2.47}
\ene

 Similarly, we denote  by $\H_{j H}$ the Hilbert space of all
measurable, $\CE^3-$ valued functions defined on $K_j$  that are square integrable with the weight
$\mu^{\lambda\nu}_j$ and the scalar product,
\beq
\left(\h^{(1)}_j, \h^{(2)}_j\right)_{j H}:= \int_{K_j}H^{(1)}_{j\lambda}\,\mu^{\lambda\nu}_j\,
\overline{H^{(2)}_{j\nu}}\, d\x^3.
\label{2.48}
\ene

The Hilbert space of finite energy fields in $K_j$ is the direct sum

\beq
\H_{j}:= \H_{j E}\oplus \H_{j H},
\label{2.49}
\ene
and the Hilbert space in the cloaked objects $K$ is the direct sum,

$$
\H_K:=  \oplus_{j=1}^N \H_{j}.
$$

The complete Hilbert space of finite energy fields  including the cloaked objects is,

\beq
\H:= \H_\Omega\oplus \H_K.
\label{2.50}
\ene

We now write (\ref{2.1}) as a Schr\"odinger equation in each $K_j$ as before.
We define the following formal differential operator,
\beq
a_j \left(\begin{array}{c}\e_j\\ \h_j\end{array}\right)=i \left(\begin{array}{c}
\var_j \nabla\times \h_j\\- \mu_j \nabla\times \e_j\end{array}\right).
\label{2.51}
\ene

Equation (\ref{2.1}) in $K_j$ is equivalent to
\beq
i\frac{\partial}{\partial t}\left(\begin{array}{c}\e_j\\\h_j\end{array}\right)= a_j\left(
\begin{array}{c}\e_j\\\h_j\end{array}\right).
\label{2.52}
\ene
Let us denote the interior of
$K_j$  by $\stackrel{o}K_j:= K_j \setminus \partial K_j$.
 Then, $a_j$ with domain $C^1_0(\stackrel{o}K_j)$ is a symmetric operator in $\H_{j}$.
We denote,

\beq
a:= a_\Omega\oplus a_K, \, \hbox{\rm where}\, a_K:= \oplus_{j=1}^N a_j,
\label{2.53}
\ene
with domain,
\beq
D(a):= \left\{ \left(\begin{array}{c}\e_\Omega\\ \h_\Omega\end{array}\right)\oplus_{j=1}^N
\left(\begin{array}{c}\e_j\\ \h_j\end{array}\right)
\in \C^1_0(\Omega)\oplus_{j=0}^N \C^1_0(\stackrel{o}K_j)\right\}.
\label{2.54}
\ene
The operator $a$ is symmetric in $\H$. The possible unitary dynamics that preserve energy for the whole
system, including the cloaked objects, $K$, are given by the self-adjoint extensions of $a$.
We have that,

\begin{theorem} \label{th-2.5}
Every self-adjoint extension, $A$, of $a$ is the direct sum of  $A_\Omega$ and of some self-adjoint
extension, $A_K$, of $a_K$, i.e.,
\beq
A= A_\Omega \oplus A_K.
\label{2.55}
\ene
\end{theorem}
This theorem tells us that the cloaked objects $K$ and the exterior $\Omega$ are completely decoupled and
that we are free to choose any boundary
condition inside the cloaked objects $K$ that makes $a_K$ self-adjoint without disturbing the cloaking
effect in $\Omega$. Boundary conditions that make
$A_K$ self-adjoint are well known. See for example, \cite{bs}, \cite{lei}, \cite{pi1} and \cite{pi2}.

It follows from explicit computation that zero is an eigenvalue of every $A_K$ with infinite multiplicity
 and that,

\beq \H_{K \perp}:= \left(\hbox{kernel}\, A_K\right)^\perp \subset
\left\{\left(\begin{array}{c}\e\\\h\end{array} \right)\in \H_K:
\frac{\partial}{\partial x_\lambda} \var^{\lambda\nu}_K E_{\nu}=0 ,
\frac{\partial}{\partial x_\lambda} \mu^{\lambda\nu}_K
H_{\nu}=0\right\}, \label{2.56} \ene where  we denote by $
\var^{\lambda\nu}_K(\x):= \var^{\lambda\nu}_j(\x)$ for $\x \in K_j$,
and $ \mu^{\lambda\nu}_K(\x):= \mu^{\lambda\nu}_j(\x)$ for $\x \in
K_j, j=1,2,\cdots,N$. It follows that zero is an eigenvalue of $A$
with infinite multiplicity and that,
 \beq \H_{ \perp}:=
\left(\hbox{kernel}\, A\right)^\perp = \H_{\Omega \perp}\oplus \H_{K
\perp}. \label{2.57} \ene For any $\varphi=
\varphi_\Omega\oplus\varphi_K \in \H_{ \perp} \cap D(A)$,

\beq
e^{-itA}\varphi= e^{-it A_\Omega}\, \varphi_\Omega \oplus e^{-itA_K}\,\varphi_K
\label{2.58}
\ene
is the unique solution of Maxwell's equations (\ref{2.1}, \ref{2.2}) with finite energy that is equal
to $\varphi$ at $t=0$. This shows once again that the dynamics in $\Omega$ and in $K$ are completely
 decoupled. If at $t=0$ the electromagnetic fields are zero in $\Omega$, they remain equal to zero
for all times, and viceversa. Actually, electromagnetic waves inside  the cloaked objects are not allowed
to leave them, and viceversa, electromagnetic waves
outside can not go inside. In general, the solutions will be discontinuous at $\partial K$.
This implies, in particular, that the presence of active devices inside the cloaked objects  has no effect
on the cloaking outside.
In terms of boundary conditions this means that transmission conditions that link the electromagnetic
fields
inside and outside the cloaked objects are not allowed.
Furthermore, choosing a particular self-adjoint extension of the electromagnetic propagator of
the cloaked objects  can be understood as  choosing
some boundary condition on the inside of the boundary of the cloaked objects.
In other words, any possible unitary dynamics that preserves energy implies the existence of some
self-adjoint extension in $K$, that can be understood as a boundary condition   on the inside of the
boundary of the cloaked objects.
The particular self-adjoint extension, or boundary condition, that nature  will take  depends on the
specific
properties of the
metamaterials used to build the transformation media as well as on the properties of the media inside the
cloaked objects. Note that this does not mean that we have to put any physical surface, a lining, on the
surface of the cloaked objects to enforce any particular boundary condition  on the inside, since as we
 already
mentioned this plays no role in  the cloaking outside. It would be, however, of theoretical interest to see
what the self-adjoint extension in $K$, or the interior boundary condition, turns out to be for specific
cloaked objects and metamaterials.
These results apply to the  exact transformation media that we consider on this paper.
However, the fact that there is a large class of self-adjoint extensions -or boundary conditions- that can be
taken inside the cloaked objects could be useful in order to enhance cloaking in practice, where one has to
consider  approximate transformation media,  as well as in the analysis of the stability of cloaking.

\section{Scattering at  Fixed Frequency}
\sss

We now consider the scattering of plane waves by the cloaks. We have the following result.
\begin{theorem}\label{th-3.1}
For any incident plane wave at any frequency the scattered wave is identically zero.
\end{theorem}
\noindent{\it Proof:} For simplicity, let us first consider the case where the $\var_0^{\lambda\nu},
\mu_0^{\lambda\nu}$ are isotropic, i.e., $\var_0^{\lambda\nu} = \var^0 \delta^{\lambda\nu},
\mu_0^{\lambda\nu}= \mu^0 \delta^{\lambda\nu}$ and that we have an incident field that propagates along
 the vertical axis, $x_3$, with the electric field polarized along the $x_1$ direction,

$$
\left(\begin{array}{c} \e^{\i}\\ \h^{\i}\end{array}\right)(\x)= \left(\begin{array}{c}{\mathbf e}_1\,
e^{i (kx^3- \omega t)}
\\\\ \sqrt{\frac{\var_0}{\mu_0}} \,{\mathbf e}_2 \, e^{i\,(k x^3-\omega t)}\end{array}\right),
$$
where $ k= \omega\sqrt{\var^0 \mu^0}$, and $ {\mathbf e}_1 , {\mathbf e}_2$ are  unit vectors, respectively,
 in the $x^1$ and $x^2$ directions.

We have to find the solution with locally finite energy to (\ref{2.3}, \ref{2.4}), $(\e,\h)^T$, that
outside the cloaks is equal to the sum
of the incident plane wave and the scattered wave, i.e.,
\beq
\left(\begin{array}{c} \e\\ \h\end{array}\right)(\x)= \left(\begin{array}{c} \e^{\i}\\ \h^{\i}\end{array}\right)
(\x)+
\left(\begin{array}{c} \e^{\s}\\ \h^{\s}\end{array}\right)(\x), \x \in \ \ere^3 \setminus
\cup_{j=1}^N B_{b_j}(\c_j).
\label{3.1}
\ene
Furthermore, the scattered wave $(\e^{\s},\h^{\s})^T$ has to satisfy the outgoing Silver-M\"uller
radiation condition.
See, for example, \cite{vb}. We compute this solution in a simple way using  the unitary   equivalence
of $A_\Omega$
 and $A_0$. By  (\ref{2.43}) the fields with locally finite energy,

$$
\left(\begin{array}{c} \e\\ \h\end{array}\right)(\x):=
\left( U  \left(\begin{array}{c}{\mathbf e}_1\, e^{i (ky^3- \omega t) }
\\ \\\sqrt{\frac{\var_0}{\mu_0}} \,{\mathbf e}_2 \, e^{i\,(k y^3-\omega t) }
\end{array}\right)\right)(\x),
$$
are a solution to (\ref{2.3}, \ref{2.4}). Furthermore,
since for $\x \in \ere^3 \setminus \cup_{j=1}^N B_{b_j}(\c_j),
A^\lambda_{\lambda'}= \delta^{\lambda}_ {\lambda'}$, we have that,
$$
\left(\begin{array}{c} \e\\ \h\end{array}\right)(\x)=  \left(\begin{array}{c}{\mathbf e}_1\,
e^{i (kx^3- \omega t) }
\\ \\\sqrt{\frac{\var_0}{\mu_0}} \,{\mathbf e}_2 \, e^{i\,(k x^3-\omega t) }\end{array}\right),\,
\x \in \ere^3
\setminus \cup_{j=1}^N B_{b_j}(\c_j),
$$
what proves that the scattered wave is identically zero.

Let us now consider the case where $ \var^{\lambda\nu}_0, \mu^{\lambda\nu}_0$ are general anisotropic media.
For a discussion of plane wave solutions in this case see, for example, \cite{we1,we2,wi}. A general incident
plane wave is given by,

$$
\left(\begin{array}{c} \e^{\i}\\ \h^{\i}\end{array}\right)(\x)= \left(\begin{array}{c} \e^{\i}({\mathbf k})\,
e^{i({\mathbf k}\cdot \x-\omega t)}
\\ \h^{\i}({\mathbf k}) \, e^{i({\mathbf k}\cdot \x-\omega t) }\end{array}\right),
$$

where $ {\mathbf k}\in \ere^3, {\mathbf k}\neq 0$ and,
$$
\left({\mathbf k} \times \e^{\i}({\mathbf k})\right)^\lambda =  \omega \mu^{\lambda \nu} \h^{\i}_\nu({\mathbf k})
 , \quad
 \left({\mathbf k} \times \h^{\i}({\mathbf k})\right)^\lambda \,=\,-\omega \var^{\lambda\nu}
 \e^{\i}_\nu({\mathbf k}).
 $$
We compute again the solution with locally finite energy using the unitary equivalence between $A_\Omega$
and $A_0$  (\ref{2.43}). We have that,

$$
\left(\begin{array}{c} \e \\ \h\end{array}\right)(\x):=
\left( U  \left(\begin{array}{c} \e^{\i}({\mathbf k})\,
e^{i({\mathbf k}\cdot \y-\omega t)}
\\ \h^{\i}({\mathbf k}) \, e^{i({\mathbf k}\cdot \y-\omega t) }\end{array}\right)\right)(\x),
$$
 are a solution to (\ref{2.3}, \ref{2.4}) with locally finite energy. As in the isotropic case,

$$
\left(\begin{array}{c} \e \\ \h\end{array}\right)(\x)=   \left(\begin{array}{c} \e^{\i}({\mathbf k})\,
e^{i({\mathbf k}\cdot \x-\omega t)}
\\ \h^{\i}({\mathbf k}) \, e^{i({\mathbf k}\cdot \x-\omega t) }\end{array}\right),\, \x \in \ere^3
\setminus \cup_{j=1}^N B_{b_j}(\c_j),
$$
what proves that the scattered wave is also zero in the general anisotropic case.

\bull

This theorem proves that we can not detect the cloaked objects $K$ in any scattering experiment.
\section{Scattering of Wave Packets}\sss
In this section we consider the scattering of finite energy wave packets. Of course, in practice one
always sends a finite energy wave packet on the target, that has to have a narrow enough
range of frequencies in order that we can neglect dispersion, as we mentioned in the Introduction.

Let $\chi_\Omega$ be the characteristic function of $\Omega$, i.e., $\chi_\Omega(\x)=1
, \x \in \Omega, \chi_\Omega(\x)=0, \x \in \ere^3\setminus \Omega$.
We define,

\beq
\left(J \left(\begin{array}{c}\e_0\\ \h_0\end{array}\right)\right)(\x):=
\chi_\Omega(\x) \left(\begin{array}{c}\e_0\\ \h_0\end{array}\right)(\x).
\label{4.1}
\ene

By (\ref{5.3}) $J$ is a bounded operator from $\H_0$ into $\H_\Omega$.

The wave operators are defined as follows,
\beq
W_\pm = \hbox{\rm s-}\lim
_{t \rightarrow \pm \infty} e^{itA_\Omega}\, J e^{-itA_0} P_{0\perp},
\label{4.2}
\ene
provided that the strong limits exist, and where $P_{0\perp}$ denotes the projector onto $\H_{0\perp}$.

Let us designate by $\mathbf W^{1,2}(\cl)$ the Sobolev space of $\mathbf \CE^6$ valued functions.
We denote by $I$ the identity operator on $\H_0$.
Then,
\begin{lemma}\label{lemm-4.1} The wave operators (\ref{4.2}) exist and,
\beq
W_{\pm}= U P_{0\perp}.
\label{4.3}
\ene
\end{lemma}
\noindent {\it Proof:} Denote,

$$
W(t):= e^{itA_\Omega}\, J \, e^{-itA_0} P_{0\perp}.
$$
By (\ref{2.43}), for any $\varphi \in \H_0$
\beq
W(t)\varphi = \psi(t)+ U P_{0\perp}\varphi,
\label{4.4}
\ene
with
\beq
\psi(t):= U \,e^{itA_0}\, \left(U^\ast J-I \right)\, e^{-itA_0} P_{0\perp}\varphi.
\label{4.5}
\ene

Since for $|\y|\geq R$, with $R$ large
 enough, our transformation, $ \x=f(\y)$,
 is the identity,
$\x=\y$, and in consequence, $A^\lambda_{\lambda'}(\y)= \delta^{\lambda}_{\lambda'}$ for $|\y| \geq R$, we
have that,

\beq
\left(U^\ast J -I\right)= \left(U^\ast J -I\right) \chi_{ B_R(0)}.
\label{4.6}
\ene
It follows that,

\beq
\hbox{\rm s-}\lim_{t \rightarrow \pm \infty} \psi(t)= U\, \hbox{\rm s-}\lim_{t \rightarrow \pm \infty} e^{itA_0}
\vartheta(t)
\label{4.7}
\ene
with,
\beq
\vartheta(t):=  \left(U^\ast J -I\right) \chi_{ B_R(0)} e^{-itA_0} P_{0\perp} \varphi.
\label{4.8}
\ene

We have that,

\beq
\begin{array}{c}
\left\|\vartheta(t)\right\|_{\H_0}  \leq  \left\| J \chi_{ B_R(0)} e^{-itA_0} P_{0\perp}\varphi\right\|_{\H}
+ \left\|\chi_{ B_R(0)} e^{-itA_0} P_{0\perp}\varphi\right\|_{\H_0} \\ \\
\leq C  \left\|\chi_{ B_R(0)}
e^{-itA_0}
P_{0\perp}\varphi\right\|_{\H_0}.
\end{array}
\label{4.9}
\ene

Then, as  $(A_0+i)^{-1} P_{0\perp}$
is bounded from $\H_0$ into $\mathbf W^{1,2}(\cl)$  \cite{we1} \cite{we2}, it follows from the Rellich
local compactness theorem that

$$
\chi_{ B_R(0)} \, (A_0+i)^{-1} P_{0\perp}
$$
is a compact operator in $\H_0$.  Suppose that $ \varphi \in D(A_0) \cap \H_{0\perp}$. Then,

\beq
\hbox{\rm s-} \lim_{t \rightarrow \pm \infty} \chi_{ B_R(0)} e^{-itA_0} P_{0\perp}\varphi =
\hbox{\rm s-} \lim_{t \rightarrow \pm \infty} \chi_{ B_R(0)}   (A_0+i)^{-1} P_{0\perp}  e^{-itA_0} (A_0+i)
 \varphi=0,
\label{4.10}
\ene
and whence, by (\ref{4.9}),
\beq
\hbox{\rm s-}\lim_{t \rightarrow \pm \infty} \vartheta(t)= 0,
\label{4.11}
\ene
and it follows that in this case,

\beq
  \hbox{\rm s-}\lim_{t \rightarrow \pm \infty}    \psi(t)=0.
\label{4.12}
\ene
By continuity this is also true for $\varphi \in \H_{0\perp}$.

Then, (\ref{4.3}) follows from (\ref{4.4}) and (\ref{4.12}).

\bull

The scattering operator is defined as

\beq
S:= W_+^\ast\, W_-.
\label{4.13}
\ene
\begin{corollary}\label{cor-4.2}
\beq
S=P_{0\perp}.
\label{4.14}
\ene
\end{corollary}
\noindent {\it Proof:} This is immediate from (\ref{4.3}) because $U^\ast\, U=I$.

\bull

Let us denote by $S_\perp$ the restriction of $S$ to $\H_{0\perp}$. $S_{\perp}$ is the physically
relevant
scattering operator that acts in the
Hilbert space $\H_{0\perp}$ of finite energy fields that satisfy equations (\ref{2.2}). We designate by
$I_\perp$ the identity operator on
$\H_{0\perp}$. We have that,

\begin{corollary} \label{cor-4.3}
\beq
S_\perp= I_\perp.
\label{4.15}
\ene
\end{corollary}
\noindent {\it Proof:} This follows from Corollary \ref{cor-4.2}.

\bull

The fact that $S_\perp$ is the identity operator on $\H_{0\perp}$  means that there is cloaking for
all frequencies. Suppose that for
 very
negative times we are given an incoming wave packet
$e^{-it A_0}\varphi_-$, with
$\varphi_- \in \H_{0\perp}$. Then, for large positive times the outgoing wave packet is given by
 $e^{-it A_0}\varphi_+$ with
$\varphi_+= S_\perp\varphi_-$.
But, as $S_\perp=I$, we have that $\varphi_+= \varphi_-$ and then,

$$
e^{-it A_0}\varphi_- = e^{-it A_0}\varphi_+.
$$

Since the incoming and the outgoing wave packets are the same there is no way to detect the cloaked
objects $K$ from scattering experiments
performed in
$\Omega$.

\section{The Proofs of Theorems \ref{th-2.3} and \ref{th-2.5}}
\sss
It follows from  (\ref{2.8}) that the transformation matrix (\ref{2.12}) is given by,

\beq
\begin{array}{c}
 A^\lambda_{\lambda'}= \frac{g_j(|\y-\c_j|)}{|\y-\c_j|} \delta^\lambda_{\lambda'}+
 \left( \frac{g_j'(|\y-\c_j|)}{|\y-\c_j|^2}-\frac{g_j(|\y-\c_j|)}{ |\y-\c_j|^3}\right)
 (\y-\c_j)^\lambda (\y-\c_j)^{\lambda'},\\\\
 \y \in B_{b_j }({\mathbf c}_j),
1 \leq j \leq N, \\\\
  A^\lambda_{\lambda'} = \delta^\lambda_{\lambda'}\ , \y \in  \cl
  \setminus \cup_{j=1}^N B_{b_j }({\mathbf c}_j).
\end{array}
\label{5.1}
\ene
The determinant is equal to,
\beq
\begin{array}{c}
\Delta(\y)= g_j'(|\y-\c_j|) \left( \frac{g_j(|\y-\c_j|)}{|\y-\c_j|}\right)^2,
 \y \in B_{b_j }({\mathbf c}_j),
1 \leq j \leq N, \\\\
  \Delta(\y)=1 , \y \in  \cl
  \setminus \cup_{j=1}^N B_{b_j }({\mathbf c}_j).
\end{array}
\label{5.2}
\ene
This result is easily obtained rotating into a coordinate system such that, $\y-\c_j=( |\y-\c_j|,0,0)$
\cite{sps}. Then, by (\ref{2.19})  the matrices $ \var^{\lambda \nu}, \mu^{\lambda \nu}$ are degenerate at
 $\partial K$ and by (\ref{2.20}) the matrices  $ \var_{\lambda \nu}, \mu_{\lambda \nu}$ are singular at
 $\partial K$.
Moreover, by (\ref{2.17}, \ref{5.1}, \ref{5.2}) there is a constant $C$ such that,

\beq
\left| \var^{\lambda \nu}(\x)\right| \leq  C,\, \, \left| \mu^{\lambda \nu}(\x)\right| \leq  C, \x \in
\Omega.
\label{5.3}
\ene

\noindent {\it Proof of theorem \ref{th-2.3}}:
Let us denote,
$$
\mathcal Q:= \left\{ (\e_o, \h_0) \in \H_0: (\e_0,\h_0)= U^\ast(\e,\h ) \,\hbox{\rm for some}\, (\e,\h)\in
C^1_0 (\Omega)\right \},
$$
where $U$ is defined in (\ref{2.42}). Let us prove that $ \mathcal Q \subset D(A_0)$. Note that
$ \mathcal Q \subset \C^1(\Omega_0 \setminus \cup_{j=1}^N \partial B_{b_j}(\c_j))$. Suppose that
$(\e_0, \h_0)  \in {\mathcal Q}$. Then,
$$
a_0 \left(\begin{array}{c}\e_0 \\\h_0\end{array}\right) \in \C\left(\Omega_0 \setminus
\cup_{j=1}^N \partial B_{b_j}(\c_j)\right),
$$
where the derivatives are defined in classical sense on $ \Omega_0 \setminus \cup_{j=1}^N
\partial B_{b_j}(\c_j)$.
Furthermore,  by the invariance of Maxwell's equations,
\beq
a_0 \left(\begin{array}{c}\e_0 \\\h_0\end{array}\right)(\y)= U^{\ast} a_\Omega \, U
\left(\begin{array}{c}\e_0 \\\h_0\end{array}\right)(\y), \, \y \in \Omega_0 \setminus \cup_{j=1}^N \partial
 B_{b_j}(\c_j).
\label{5.4}
\ene
This implies that,

\beq
a_0 \left(\begin{array}{c}\e_0 \\\h_0\end{array}\right)\in \H_0.
\label{5.5}
\ene
But, as $(\e_0,\h_0)$ have continuous tangential components at $\cup_{j=1}^N \partial
 B_{b_j}(\c_j)$  the function in the left-hand side of
(\ref{5.4})  defined  for $\y \in \Omega_0 \setminus
\cup_{j=1}^N \partial B_{b_j}(\c_j)$   with the derivatives in classical sense  actually coincides with
the distribution that is obtained when the derivatives are taken in distribution sense in $\cl$. Then,
by (\ref{5.5})   we have that  $(\e_0,\h_0)\in D(A_0)$, and

$$
A_0 \left(\begin{array}{c}\e_0 \\\h_0\end{array}\right) = a_0 \left(\begin{array}{c}\e_0 \\\h_0
\end{array}\right)
$$
with the right-hand side defined as indicated above. By (\ref{5.4}) this implies that,
\beq
A_0|_{\mathcal Q}= U^\ast a_\Omega \, U,
\label{5.6}
\ene
or what is equivalent, that
\beq
a_\Omega= U A_0|_{\mathcal Q}\, U^\ast.
\label{5.7}
\ene
It follows that to prove that $a_\Omega$ is essentially self-adjoint and that (\ref{2.43}) holds
we have to prove that $A_0$ is essentially self-adjoint in $\mathcal Q$.

In the proof below we denote by $a_\Omega$ the formal differential operator.

Suppose  that $ (\e_0, \h_0)\in C^1_0(\Omega_0)$. Then,
\beq
\left(\begin{array}{c}\e_0 \\\h_0\end{array}\right) = U^\ast \left(\begin{array}{c}\e \\\h\end{array}\right),
\, \hbox{\rm with}\,   \left(\begin{array}{c}\e \\\h\end{array}\right)= U \left(\begin{array}{c}\e_0 \\\h_0
\end{array}\right).
\label{5.8}
\ene
Hence, arguing as above, but in the opposite direction,
we prove that,

\beq
a_\Omega \left(\begin{array}{c}\e \\\h\end{array}\right)= U a_0
\left(\begin{array}{c}\e_0 \\\h_0\end{array}\right) \in \H_\Omega,
\label{5.9}
\ene
where the function in the left-hand side with the derivatives taken
in classical sense in $ \Omega \setminus
\cup_{j=1}^N \partial B_{b_j}(\c_j)$ actually coincides with the distribution obtained when the
derivatives are taken
in distribution sense in $\Omega$. Let us now introduce a mollifier. Take $ f \in C^\infty_0(\ere^3),
f(\x)=1, |\x| \leq 1/2, f(\x)=0, |\x| \geq 1, \int f(x)dx =1$, and define $f_n
(x)= n^3 f(nx)$. Denote,
\beq
\left(\begin{array}{c}\e \\\h\end{array}\right)_n :=
\int_{\ere^3} \left(\begin{array}{c}\e \\\h\end{array}\right)
(\x-{\mathbf z}) f_n({\mathbf z})\, d{\mathbf z}\in C^\infty_0(\Omega),
\label{5.10}
\ene
for $n$ large enough, where we have extended $(\e,\h)^T$ to $\ere^3$ by zero. Moreover,
\beq
\left(\begin{array}{c}\e \\\h\end{array}\right) = \hbox{\rm s-}\lim_{n \rightarrow \infty}
 \left(\begin{array}{c}\e \\\h\end{array}\right)_n,
\label{5.11}
\ene
and

\beq
 \hbox{\rm s-}\lim_{n \rightarrow \infty}
a_\Omega\left(\begin{array}{c}\e \\\h\end{array}\right)_n =   \hbox{\rm s-}\lim_{n \rightarrow \infty}
 \int_{\ere^3} a_\Omega\left(\begin{array}{c}\e \\\h\end{array}\right)
(\x-{\mathbf z}) f_n({\mathbf z})\, d{\mathbf z} =  a_\Omega\left(\begin{array}{c}\e \\\h
\end{array}\right)
(\x).
\label{5.12}
\ene
The limits in (\ref{5.11}, \ref{5.12}) are in the strong convergence in $\H_\Omega$.
It follows from (\ref{5.6}, \ref{5.8}, \ref{5.11}, \ref{5.12}) that $(\e_0, \h_0)^T$ can be approximated in the
graph norm of $A_0$ by functions in $\mathcal Q$. Then, it is enough to prove that $A_0$ is essentially
self-adjoint in $C^1_0(\Omega_0)$.
But as $A_0$ is essentially self-adjoint in $C^\infty_0(\cl)$ it is also essentially self-adjoint in
 $C^1_0(\cl)$. Hence, it is enough to prove that any
function in $C^1_0(\cl)$ can be approximated  in the graph norm of $A_0$ by functions in
$C^1_0(\Omega_0)$. To prove this take any
continuously differentiable real-valued
function, $f$,
defined on $\cl $ such that, $f(y)=0, |y|\leq 1$ and $f(y)=1, |y| \geq 2$. Then, for any

$$
\left(\begin{array}{c}\e_0\\\h_0\end{array}\right) \in \C^1_0(\cl),
$$
we have that,
$$
     \prod_{j=1}^ N f(n|\y-\c_j|)\,
     \left(\begin{array}{c}\e_0\\\h_0\end{array}\right) \in
     \C^1_0(\Omega_0),
$$
and moreover,
$$
\hbox{\rm s-}\lim_{n\rightarrow \infty}  \prod_{j=1}^ N f(n|\y-\c_j|)\, \left(\begin{array}{c}
\e_0\\\h_0\end{array}\right)=  \left(\begin{array}{c}\e_0\\\h_0\end{array}\right),
$$
$$
\hbox{\rm s-}\lim_{n\rightarrow \infty} a_0  \prod_{j=1}^ N f(n|\y-\c_j|)  \, \left(\begin{array}{c}
\e_0\\\h_0\end{array}\right)=
a_0 \left(\begin{array}{c}\e_0\\\h_0\end{array}\right),
$$
where by $\hbox{\rm s-}\lim$ we designate the strong limit in $\H_0$. This completes the proof
that $a_\Omega$ with domain $C^1_0(\Omega)$ is essentially self-adjoint and that (\ref{2.43}) holds.

The unitary equivalence given by (\ref{2.43}) implies that $A_\Omega$ has the same spectral properties
that $A_0$. Namely, it has no
singular-continuous spectrum, the absolutely-continuous spectrum is $\ere$ and the only eigenvalue is
 zero and it has infinite multiplicity.
Moreover,

\beq
\H_{\Omega \perp}:= \left(\hbox{kernel}\, A_\Omega\right)^\perp = \left\{\left(\begin{array}{c}\e\\
\h\end{array}\right)\in \H_\Omega:
\frac{\partial}{\partial x_\lambda}
\var^{\lambda\nu}E_{\nu}=0 ,  \frac{\partial}{\partial x_\lambda}
\mu^{\lambda\nu}H_{\nu}=0\right\}.
\label{5.13}
\ene

\bull

\noindent{\it Proof of  Theorem \ref{th-2.5}:}
Let us
denote by $\overline{a}$ the closure of $a$, with similar notation for $a_\Omega,
a_j, j=1,\cdots,N$. Then,
$$
\overline{a}= A_\Omega \oplus_{j=1}^N \overline{a_j},
$$
where we used the fact that as $a_\Omega$ is essentially self-adjoint, $\overline{a_\Omega}=A_\Omega
$.
The adjoint of $a$ is given by,

\beq
D(a^\ast)= \left\{ \left(\begin{array}{c}\e_\Omega\\ \h_\Omega\end{array}\right)\oplus_{j=1}^N
\left(\begin{array}{c}\e_j\\ \h_j\end{array}\right)
\in \H :    \left(\begin{array}{c}\e_\Omega\\ \h_\Omega\end{array}\right) \in D(A_\Omega), a_j
\left(\begin{array}{c}\e_j\\ \h_j\end{array}\right)
\in \H_j \right\},
\label{}
\ene
and
\beq
a^\ast \left( \left(\begin{array}{c}\e_\Omega\\ \h_\Omega\end{array}\right)\oplus_{j=1}^N
\left(\begin{array}{c}\e_j\\ \h_j\end{array}\right)\right)
= A_\Omega \left(\begin{array}{c}\e_\Omega\\ \h_\Omega\end{array}\right) \oplus_{j=1}^N a_j
\left(\begin{array}{c}\e_j\\ \h_j\end{array}\right),
\label{5.14}
\ene
for
\beq
\left(\begin{array}{c}\e_\Omega\\ \h_\Omega\end{array}\right)\oplus_{j=1}^N \left(\begin{array}{c}
\e_j\\ \h_j\end{array} \right)\in
D(a^\ast),
\label{5.15}
\ene
where the derivatives are taken in distribution sense.

Let us denote by ${\mathcal K}_{\Omega \pm}:= \hbox{kernel}(i  \mp a_\Omega^\ast),
{\mathcal K}_{j \pm}:= \hbox{kernel}(i  \mp a_j^\ast)$
the deficiency subspaces of $ a_\Omega$ and $a_j, j=1,\cdots,N$. Since $a_\Omega$ is
essentially self-adjoint
$\mathcal K_{\Omega \pm} =\{0\}$. Let  $\mathcal K_\pm:= \oplus_{j=1}^N \mathcal K_{j\pm}$
be the deficiency subspaces of
$ a_K:=\oplus_{j=1}^N a_j$. Suppose that $\mathcal K_\pm$ have the same dimension. Then, it
follows from Corollary 1 in page 141 of \cite{rsII}
that there is a one-to-one correspondence between   self-adjoint extensions
of $a_K$ and unitary maps
from $\mathcal K_+$  into $\mathcal K_-$. If $V$ is such a unitary map, then, the corresponding
self-adjoint extension $A_{KV}$ is given by,
$$
D(A_{K V})=\left\{ \varphi +\varphi_+ +V\varphi_+ : \varphi \in D(\overline{a_K}), \varphi_+
 \in \mathcal K_+\right\},
$$
and
$$
A_{KV} \varphi = \overline{a_K}\varphi+ i\varphi_+ -i V\varphi_+.
$$
Hence, since $\mathcal K_{\Omega \pm} =\{0\}$ and $\overline{a}= A_\Omega \oplus \overline{a_K}$
there is a one-to-one correspondence  between
self-adjoint extensions
of $a$ and unitary maps, $V$, from $\mathcal K_+$  into $\mathcal K_-$. The self-adjoint extension
 $A_V$  corresponding to $V$ is given by,
$$
A_V =A_\Omega\oplus A_{KV}.
$$

\section{Generalizations}\sss
In this paper we considered  high order transformation media obtained from  singular transformations
 that blow up a finite number of points, by simplicity,
and since this is the situation in the applications. Suppose that we have a transformation that is singular
in a set of points that we  call
$M$ and denote now $\Omega_0:= \cl \setminus M$. What we really used in the proofs is that
$\mathbf W^{1,2}(\cl)= \mathbf W^{1,2}_0(\Omega_0)$ where $\mathbf W^{1,2}_{0}(\Omega_0)$ denotes the
completion of $\mathbf C^\infty_0(\Omega_0)$ in the norm of $\mathbf W^{1,2}(\cl)$. We also assumed
that $\var_0^{\lambda\nu}, \mu^{\lambda\nu}_0$ are constant. What was actually  needed  is that $a_0$ is
essentially self-adjoint. Our results
hold under this more general conditions provided that in (\ref{4.2}, \ref{4.3}) and (\ref{4.14}) we replace
$P_{0\perp}$
by the projector onto the absolutely-continuous subspace of $A_0$ and that we assume that
$D(A_0) \cap \H_{0ac} \subset \mathbf W^{1,2}(\cl)$, where we
have denoted the absolutely-continuous
subspace of $A_0$ by  $ \H_{0ac}$. Moreover, $S_\perp$ has to be defined as the restriction of
$S$ to $\H_{0ac}$ and in (\ref{4.15}) $I_\perp$ has to
be the identity operator on $\H_{0ac}$. Note that under these general assumptions $A_0$ could have non-zero
eigenvalues and singular-continuous
spectrum.

For example, $\mathbf W^{1,2}(\cl)= \mathbf W^{1,2}_0(\Omega_0)$ if $M$   has zero Sobolev one capacity
\cite{af,kkm, km}.
Moreover, assume that the permittivity and the permeability tensor densities $\var_0^{\lambda\nu},
 \mu^{\lambda\nu}_0$ are
bounded below and above. Under this condition   $a_0$ is  essentially self-adjoint.
Furthermore, let us denote by $\hat{\H}_0$ the Hilbert space of finite energy solutions defined as in
(\ref{2.32}) but with $\varepsilon_0^{\lambda\nu}= \mu_0^{\lambda\mu}= \delta^{\lambda\mu} $.
 Let $ \hat{A}_0,
\hat{\H}_{0\perp}$ be, respectively, the electromagnetic generator in $\hat{\H}_{0}$ and the orthogonal
complement of its kernel. We have that $\H_{0}$ and $\hat{\H}_{0}$ are the same set of functions with
equivalent norms.
Furthermore, $ D(A_0)= D(\hat{A}_0),\hbox{kernel}\, \hat{A}_0= \hbox{kernel}\, A_0$. Moreover,
 $(\e_0,\h_0)^T\in
\H_{0\perp}$
if and only if $ \e_{0\lambda}= \sum_{\mu =0}^3\var_{0\lambda\mu} \hat{\e}_{0\mu}, \,\h_{0\lambda}
= \sum_{\mu=0}^3\mu_{0\lambda\mu}
\hat{\h}_{0\mu}$ for some $(\hat{\e}_0,\hat{\h}_0)
\in
\hat{\H}_{0\perp}$.
As \cite{we1,we2} $D(\hat{A}_0)\cap \hat{\H}_{0\perp} \subset \mathbf W^{1,2}(\cl)$ we have
 that $D(A_0) \cap \H_{0\perp} \subset \mathbf W^{1,2}(\cl)$
if $ \varepsilon_0, \mu_0$ are bounded operators on $\mathbf W^{1,2}(\cl )$ and this is true if the
 derivatives
$\frac{\partial}{\partial y_\rho}\varepsilon_{0}, \frac{\partial}{\partial y_\rho}\mu_{0}$
are bounded operators on $\hat{\H}_0$ for $\rho =1,2,3$. Note, furthermore, that $\H_{0ac} \subset
\H_{0\perp}$.

\section{Conclusion and Outlook}
We gave a rigorous mathematical proof, in the time and frequency domains, that first and high order
electromagnetic invisibility cloaks actually  cloak passive and active devices in a very strong sense.
This puts the theory of cloaking in exact transformation  media in a firm mathematical basis that will allow
us, in the next step forward, to analyze the stability of cloaking  in the approximate transformation media
that are used in the applications.

A novel aspect of our results is that, as we prove cloaking for transformations media that are obtained from
general anisotropic materials, we prove that it is possible to cloak objects contained in general
anisotropic media. A very important case of anisotropic media are crystals. Suppose that we wish to cloak an
object that is contained inside a general anisotropic medium, or a crystal, with permittivity and permeability
tensors, respectively, $\var_0^{\lambda \mu}$ and $\mu_0^{\lambda \mu}$. To cloak the object we proceed as
follows. We have to coat it with a metamaterial whose permittivity and permeability tensors are obtained using
the transformation formula (\ref{2.17}) -for both of them- putting in the right-hand side, respectively, the
permittivity and the permeability of the general anisotropic medium, or the crystal. Our theory shows
that this object with passive and active devices will be cloaked inside the general anisotropic medium, or the
crystal. As we already mentioned, this is a new result, that shows that it is not necessary to transform from
isotropic media. It is possible to transform from  general anisotropic media. This opens the way to many
interesting potential applications, not only for cloaking, but also for guiding electromagnetic waves under
quite general circumstances.

\noindent {\bf Acknowledgement}

\noindent This work was partially done while I was visiting the Institut f\"ur Theoretische Physik,
Eidgen\"ossische Techniche H\"ochschule Zurich. I thank professors Gian Michele Graf and J\"urg
Fr\"ohlich for their kind hospitality. Part of this work was done during a visit to the Department of
Mathematics and Statistics of the University of Helsinki. I thank Prof.
Lassi P\"aiv\"arinta for his kind hospitality.

\vspace{1cm}
\begin{center}
\begin{figure}
\includegraphics[bb=0 490  650  790,clip,width=15cm]{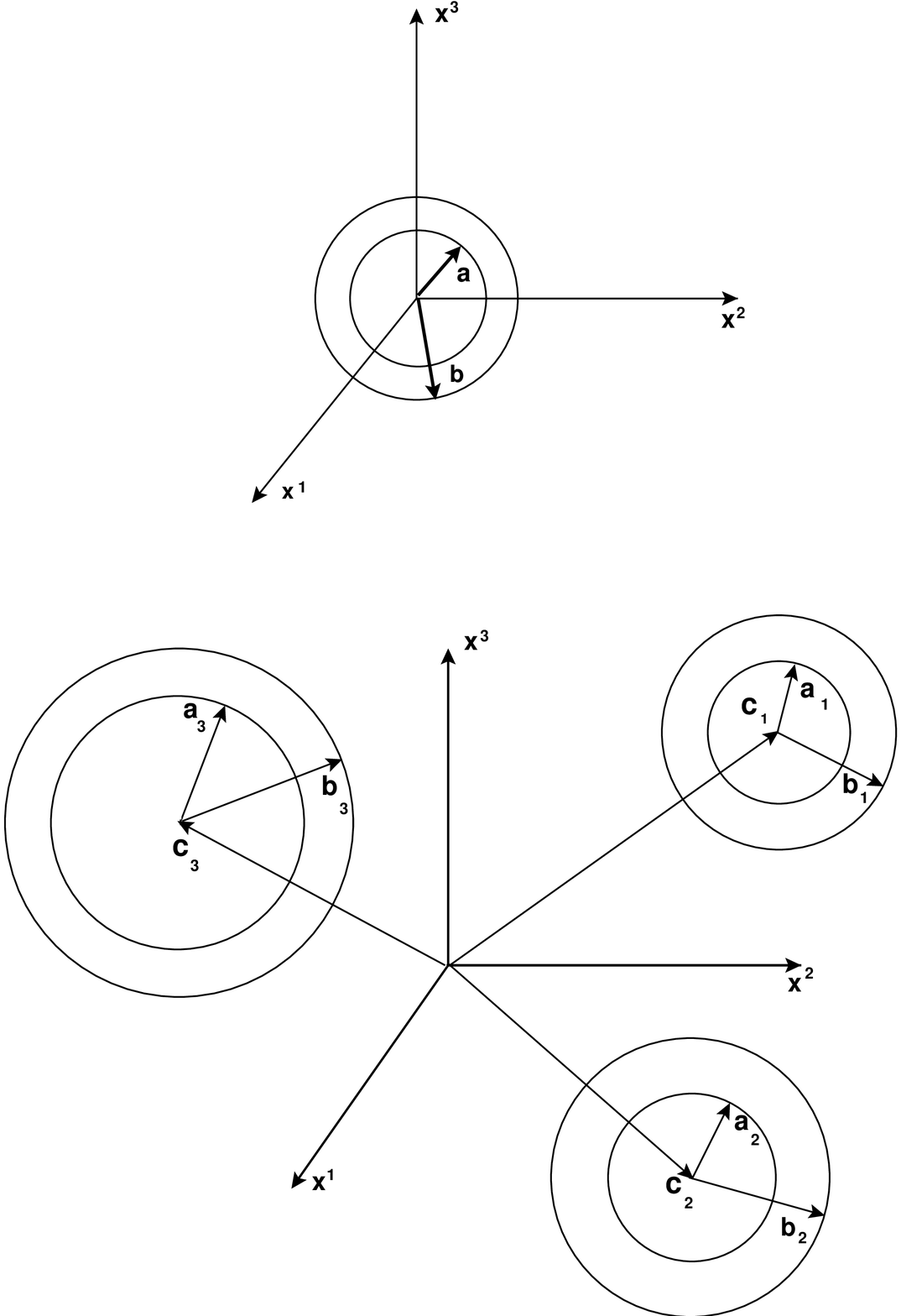}
\caption{One spherical cloak centered at zero.   }
\label{fig1}
\end{figure}

\begin{figure}
\includegraphics[bb=0 0  650  450,clip,width=15cm]{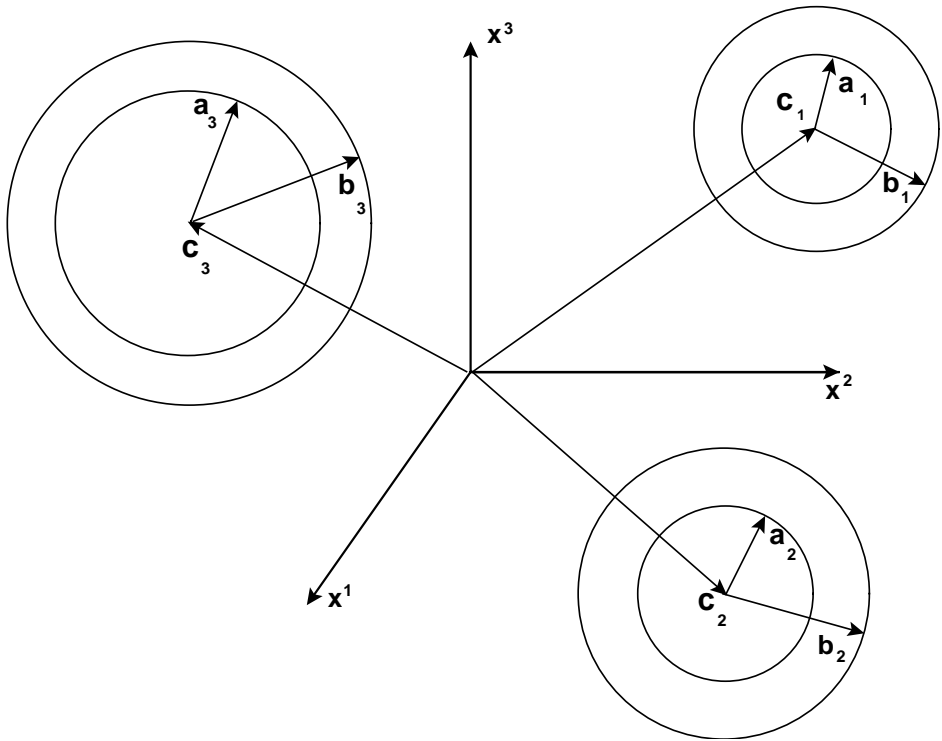}
\caption{Three spherical cloaks centered at $\c_1,\c_2,\c_3$.}
\label{fig2}
\end{figure}
\end{center}

\end{document}